\documentstyle[aps,epsfig,float,twocolumn,prl]{revtex} 
\newcommand{\be}{\begin{equation}}
\newcommand{\ee}{\end{equation}}

\begin{document}
\twocolumn[\hsize\textwidth\columnwidth\hsize\csname @twocolumnfalse\endcsname 
\draft
\title{Depinning by Fracture in a Glassy Background}
\author{M. B. Hastings, C. J. Olson Reichhardt, and C. Reichhardt}
\address{
Center for Nonlinear Studies and Theoretical Division, Los Alamos National
Laboratory, Los Alamos, New Mexico 87545
}
\date{September 23, 2002}
\maketitle
\begin{abstract}
We force a single particle through a two-dimensional simulated glass.
We find that the particle velocity obeys a
robust power law that persists to drives well above threshold.
As the single driven particle 
moves, it induces cooperative distortions in the surrounding medium.
We show theoretically that a fracture model for these 
distortions produces power law
behavior, and discuss implications for experimental probes of soft
matter systems.
\end{abstract}
]

Recently, there has been considerable interest in
the behavior of systems of interacting 
particles that resist flow, including  
vortex glasses in superconductors \cite{vg,river}, 
charge-density-wave metals\cite{cdw}, Wigner crystals \cite{Olson}, 
dense colloidal suspensions\cite{Perk,Weeks}, and
Coulomb-blockade arrays\cite{MW,cblock}.
In each of these systems,
quenched disorder and repulsive interactions from the
surrounding particles prevent any
individual particle from moving in response to a small external force.
As a result, below a threshold force, the system
behaves as a solid, while above the threshold, 
the system can flow plastically and is soft and disordered.
Once the system depins, increasing the force leads to 
rapidly increasing motion, as the speed and 
transport response increase faster 
than linearly with the flow over a range of applied force.  
Theories\cite{MW,Fisher,Nar,Watson} have attributed 
this anomalous transport to the concerted action of many driven particles 
across a rough pinscape.
They predict power-law growth of the velocity
with force in the vicinity of a threshold force $F_{c}$:
$v \sim (F-F_c)^{\beta}$.  
In experiments and simulations,
the velocity above threshold has been observed to vary as a power $\beta$ 
of the force, with 
$\beta$ in the range 1.5 to 2.2 
\cite{Olson,Perk,Shobo,Doninguez}.  
In contrast to the critical-state model of
elastic depinning\cite{Fisher}, 
where scaling occurs only very close to threshold, the power law is 
observed to hold for forces of several times the
threshold force. 

In this work we demonstrate power-law collective
transport with a {\it single} driven particle in a disordered glassy matrix 
of other non-driven particles in 
two dimensions, realized via a molecular-dynamics simulation.  
We find a power law with $\beta=1.5$ over two decades of force that is
insensitive to the system size, 
the size of the driven particle, or the density of the surrounding
medium.
This contrasts with
a single particle driven over a substrate with {\it quenched}
disorder 
where a scaling of $\beta = 1/2$ is expected \cite{Fisher}. 
Our result suggests that the origin of the
$\beta>1$ scaling in a variety of systems may be 
simpler than previously 
supposed.  A single driven particle drags other particles
with it, thus slowing it down.  The faster it moves, the fewer particles it 
drags, and hence the exponent $\beta$ becomes larger than $1.0$.  
We explicitly show that an analysis in terms of
fracture in front of the particle gives $\beta=1.5$.
The fracture leads to a one dimensional (1D) plastic zone, 
which appears as a river-like flow of particles\cite{river}.
While in other cases such rivers are
attributed to easy paths through a background of
quenched disorder, in our system the only disorder is due to the
glassiness of the medium.
The plastic zone appears due to the softness of the system;
for stiffer interactions, the plastic zone disappears and $\beta=1$ is
the observed scaling.

In addition to offering a simpler model of depinning,
our results are relevant
to systems in which two species of particles move in opposite directions
with respect to each other.
Such situations can arise in certain 
types of electrophoresis experiments, pedestrian motion,
self driven particles and molecular motors \cite{Helbing}, as well
as in vortex experiments in which a vortex lattice is driven past a low
density of pinning sites.  

Further, driving a single particle through a soft
matter system can be used as a powerful experimental probe of 
dynamics of the medium far from equilibrium.
For example, in recent experiments a magnetic particle is dragged
through a colloidal system near the glass transition\cite{Weeks}.  The
absence of momentum conservation in our system leads to very different
physics from the transfer of momentum through hydrodynamic and hard-core
interactions.  However, such experiments on colloidal systems and emulsions in 
confined geometries, in particular between parallel walls, are predicted
to have similar features to those found here.

{\it Simulations and Dissipation Balance--}
We drive a single particle with a charge $q_{D}$ at constant force
through a two dimensional (2D) disordered system with
periodic boundary conditions in the $x$ and $y$ directions.
To create a glassy medium and prevent formation of a
triangular lattice,
we simulate a mixture of two species of particles with
different charges $q_A=q$ and $q_B=2q$, where $q=1$. 
For the driven particle, we consider a range of values from
$q_{D} = 0.15q$ to $q_{D}=60q$.
The overdamped equation of motion
for particle $i$ is
$\eta {\bf v} = {\bf f}_{i} = -\sum_{j}\nabla U(r_{ij}) + F_{d}{\bf \hat{x}}$,
where ${\bf v}$ is the particle velocity, 
the damping coefficient $\eta=1$, and $F_{d}=0$ on 
all particles except the one with charge $q_{D}$.
We use a screened Coulomb interaction, given by:
\be
\label{uab}
U(r_{ij}) = q_i q_j \frac{e^{-2 r_{ij}}}{r_{ij}},
\ee
between particles $i$ and $j$ separated by a distance $r_{ij}$.
We have considered a variety of system sizes for 
$N=480$ and $N=2150$ particles, as well as 
lattice constants of $a = 1.08$ to $2.3$. 

In Fig.~1 we show the positions of the particles
and the trajectories for a fixed number of time steps 
for a system with $N=2100$ particles at a drive $F_{d} = 0.75$.
The driven particle, marked as
a large dot, has $q_{D}= 20$. 
We find that in general there is a finite threshold force $F_{c}$
for the particle to move. 
The perturbation of the other particles by the driven particle 
is anisotropic, with a larger perturbation
in the direction of drive than in the transverse direction.
Particles more than a few $a$ away from the path of the driven particle move
elastically in small nearly closed orbits of radius less than $a$.
Particles in front and behind the driven particle exhibit plastic motion.
This plastic zone {\it decreases} in size for higher velocities.  

The medium, while disordered, is a solid.  It has a threshold, $F_d=F_c$, for
failure, and can support shear stress.  Thus, the particle must {\it fracture}
the medium to move through it.  
From momentum balance, we have that at all times 
$F_d=\eta V + \eta \sum v_A$, where $V$ is the velocity of the
driven particle and the sum ranges over all other particles.  
Below threshold, where the medium moves
with the particle, the momentum balance yields $V=F_d/N$, where $N$
is the number of particles.  Above threshold, to obtain power law
scaling, some large number $n$ of other particles, $1 \ll n \ll N$, 
must move with the
driven particle, with some finite size corrections to scaling due to net 
background motion of the remaining $N-n$ particles.  Eventually, as
$n\rightarrow 1$ far above
threshold, the velocity returns to linear scaling.

In Fig.~2 
(lower solid curve) we show the log-log plot for the velocity vs applied
drive $F_{d} - F_{c}$, 
for the driven particle in Fig.~1, 
where $F_{c}=0.7$ is the threshold for the large particle to move. 
We find a good scaling with a fit of 
$\beta = 1.47 \pm 0.03$.  We find a similar scaling for 
a variety of other
parameters with the exponents of $1.5 \pm 0.05$.  At sufficiently
large drive, $\beta$ returns to 1, as expected.  
The size of the scaling region decreases with decreasing $q_D/q$,
until for $q_{D}/q < 1.0$, no anomalous scaling is observed.
In this case the driven particle
does not induce plasticity in the other particles but only a smaller
perturbation of size less than $a$. 
In the upper solid curve in Fig.~2 we show the velocity force curve for 
$q_{D} = 0.5$ showing a scaling fit with  $\beta = 1.0$.  
For a fixed
value of $F_{d}$ the velocity decreases with increasing $q_{D}$ as the
driving particle interacts more strongly with the surrounding particles. 
In Fig.~3(b), $V$ vs $q_{D}$ for $F_{d} = 4.0$ shows 
a $q_{D}^{-1/2}$ scaling in the regime where there is plastic deformation,
and then $V$ flattens out in the elastic regime for $q_D<1$. 
Additionally in the plastic 
flow regime the particle motion is highly intermittent as illustrated 
in Fig.~3(a),
where a time trace of $V$ for $q_{D}/q = 20.0$ shows the motion
occurring in bursts separated by quiet periods.  As the drive
increases, the motion becomes less intermittent, and the 
noise spectrum of the time trace develops a well-defined rollover frequency
$\omega$, with $\omega$ scaling approximately linearly in $F_d-F_c$.
We find that the rollover frequency corresponds to the time scale between peaks
in Fig.~3(a), rather than to the duration of a single peak.

In Fig.~4, we show a contour plot of the average of $v^2$ for the other 
particles in the medium, as a function of position relative to the
driven particle, for systems with $q_D/q=20$, $N=2100, a=1.085$ and
$F_d=0.9,1.5,$ and $40$ respectively.  This measures energy dissipation, as 
from energy balance 
$F_d \overline V = \eta (\overline{V^2} + \sum \overline{v_A^2})$.
These systems are above the depinning threshold of $0.7$, but
still within the scaling regime.
The dissipation is centered strongly around the moving particle,
and extends anisotropically into the surrounding medium, with a 
larger region of dissipation in the direction of drive.

{\it Theory: Elasticity and Fluid Flow---}
We now consider various scenarios for the
particle motion to understand the simulation results.  We begin by
considering the response of the medium away from the plastic zone,
where the particles move less than $a$ and elasticity theory is applicable.
The driven particle exerts a force on the
elastic medium at a location that is moving at velocity $\overline V$.
At long wavelengths, elasticity theory gives for a lattice displacement 
${\vec u}({\vec k})$ the equation of motion
\be
\label{2de}
\dot {\vec u}({\vec k})=
C_1 {\vec k} ({\vec k}\cdot {\vec u}) 
+ 
C_2 {\vec k}\times
({\vec u}\times {\vec k})
.
\ee
To obtain elastic constants $C_1,C_2$, we consider
a triangular lattice of particles, with appropriate lattice constant $a$,
and assume that all particles have the same charge, $q_{A,B}=\sqrt{2} q$.  
For a lattice constant $a=1.085$ and the chosen $q\approx 0.086$, one finds
$C_1 \approx 0.025/\eta, C_2\approx 0.0018/\eta$.
The particle exerts a force on this medium, along the direction
of motion of the particle.  There is also a force normal
to this direction, pushing the medium out sideways in opposite directions
on opposite sites of the moving fracture.  In the co-moving frame, the
displacement in any given direction decays exponentially at
large distance with a characteristic length 
$l=C_1/\overline V$ along
that direction and $l=C_2/\overline V$ normal to the direction.
The energy dissipation rate resulting is of order
$\eta \overline V^2 (l/a)^2=\eta C_1 C_2/a^2$, and is thus
independent of $\overline V$.
This indicates a force on the driven particle of order $1/\overline V$, in
addition to the drag force, $\eta \overline V$, on the driven particle.
For small $\overline V$, this force becomes arbitrarily large, ultimately
exceeding the yield stress of the medium.  In such a region the material
must yield and the elastic picture becomes invalid.

Thus, we must consider the plastic zone.  
It is not consistent to have a 2D
chunk of solid moving with the particle, with
a plastic zone between that solid and the rest of the medium: if the
particle force is sufficient to fracture the medium, it would fracture the
solid moving with it.  
At the same time, one cannot have
a 2D fluid zone
around the particle.  The overdamped nature of the dynamics (assuming
viscosity and pressure terms like a Newtonian fluid) would
concentrate all the vorticity of the fluid
within a length $a$ around the particle, and thus this scenario is also
inconsistent: more than a length $a$ from the particle, the stress is
too small to fracture the medium and produce the fluid.  By elimination
we turn next to a 1D plastic zone.

{\it Theory: Compressed Column---}
We now propose a scenario based on competition between shear
and compressive failure which accounts for all the
numerical results, and 
illustrate it using the specific systems of Fig.~4.
The driven particle creates a fracture in front of it.
Naively the fracture force would be expected to be of order the particle
interaction force $a U''(a)$, or roughly $0.015$ for the given $q_D=20$.  
In fact, it is equal to $0.7$, nearly two orders of magnitude
greater, for two reasons: 
the force $U'(r)$ rises rapidly at short $r$, while
the charge $q_D>q$ further increases the fracture force.
After fracture, 
the particles in
front of the driven particle must then move out of its way, either by failing
in shear and
moving along with the driven particle, or by failing in compression and
moving transversely out of
its way.  Initially, consider just the first possibility
so that in front of the driven particle there is a
growing 1D column of particles failing in shear.
One finds that the 
rate at which this length increases is determined {\it not} by 
the velocity of the
driven particle, but by the (faster) velocity at which a compressive front
ahead of that particle moves.
1D elasticity theory would imply that in time $t$, this compression
reaches a distance $l \approx a \sqrt{U'' t/\eta}$.  
Since this system is far from equilibrium, we have checked
this result by simulating
a 1D system with a single driven particle
in a fixed, periodic background potential to mimic the medium.
We find that this behavior 
remains valid, even close to the depinning transition, albeit with a greatly
increased value for $U''$.

The 1D column does not grow indefinitely in length due to
the possibility of transverse motion of the particles, in which case
they squirt out of the column.
The time scale for this process would naturally be of order $\eta a/(F_d-F_c)$.
Thus, the number of particles in
the column is $l/a\approx \sqrt{U'' a/(F_d-F_c)}$.   By momentum balance, this
number is equal to $\eta \overline V/(F_d-F_c)$, giving 
\be
v\propto (F_d-F_c)^{3/2},
\ee
as observed, where some fixed force $F_c$ is
required to create the fracture.  
The exit of particles from the column due to compressive failure is
naturally highly intermittent, leading to intermittent motion of
the driven particle, as seen in Fig.~3(a).  This predicts the
linear scaling of the rollover frequency with $F_d-F_c$,
with the prefactors such that the
rollover frequency is increased above the naive expectation:
at $F_d=1.5$, $\omega=0.05$, larger than $\eta a/(F_d-F_c)$. 

The 1D plastic zone is visible in
the contours shown in Fig.~4.  The aspect ratio of the contour 
increases for lower drives as
the length of the contours increases at constant width,
confirming that dissipation arises in a
1D plastic zone, rather than due to the elastic response of
a 2D medium which would instead give a constant aspect ratio.
Further, the size of the
contours is not consistent with
2D elastic response,
which would appear in Fig.~4(b) as an
exponential decay of the dissipation on a
length scale of approximately $0.35$ for $\overline V=0.072$, while 
the actual contours are significantly larger.

As the charge of the driven particle increases, it produces a depletion
zone of missing particles around it to compensate its charge; the area
of this zone is proportional to $q_D/q$.  The length of the
compression was determined above; the width will be proportional to the
radius of the depletion zone, $\sqrt{q_D/q}$, accounting for the scaling
of velocity with $q_D$, as seen in Fig.~3(b).

{\it Summary---}
We have found a robust power-law for the velocity of a single
driven particle fracturing a glassy environment, with $\beta = 1.5 \pm 0.05$. 
We give a theoretical explanation based on a
1D plastic zone, which decreases in size as the particle moves faster leading
to $\beta>1$.  This behavior arises due to the softness of the material, as for
larger lattice constants $a$ or smaller $q_D$, the interaction becomes
stiffer, and the width of the scaling region decreases until it eventually
disappears leaving only linear scaling.
While for conventional solids this scaling region is absent
with no column observed in front of the particle, the origin
of anomalous transport features in many disordered systems may be
due to the mechanisms discussed herein.

{\it Acknowledgments---}
We thank Eric Weeks for sharing his experimental results which inspired
this work.  We thank Tom Witten for many very useful discussions and
for a critical reading of the manuscript.
This work was performed at the Non-Equilibrium Summer Institute at
Los Alamos.  This work was supported by DOE Contract No. 
W-7405-ENG-36.  

\vskip-10mm
\begin{figure}
\center{
\epsfxsize=3.5in
\epsfbox{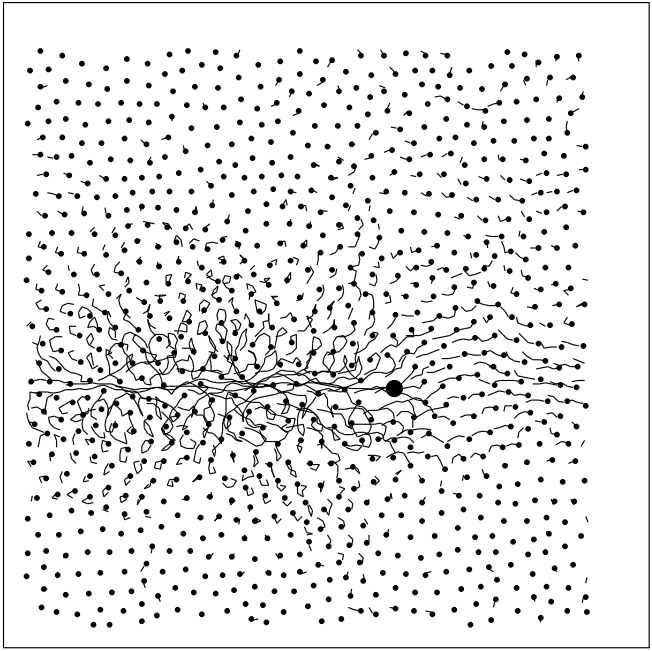}}
\caption{
The particle positions (black dots) and trajectories (black lines).
The driven particle (large dot) has charge $q_{D} = 20$; the
remaining particles are a mixture of $q_{A} = q$ and $q_{B}=2q$. 
There are a total of 2100 particles. 
The trajectories are drawn for a fixed number of time 
steps with a constant applied drive of $F_{d} = 0.75$.}
\end{figure}   

\begin{figure}
\center{
\epsfxsize=3.5in
\epsfbox{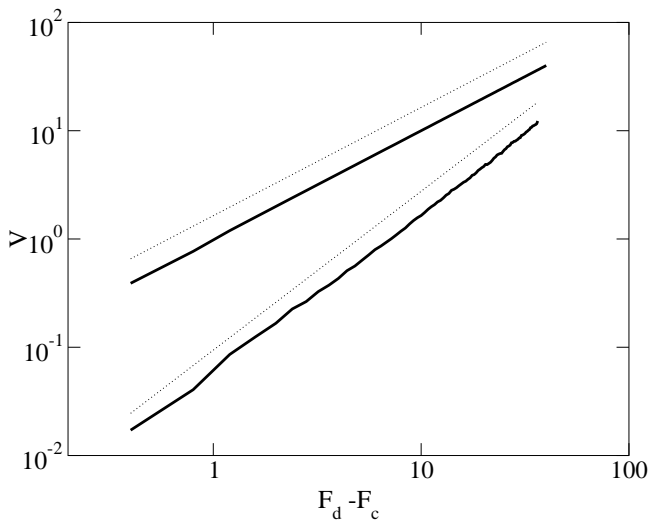}}
\caption{
The velocity $V$ vs $F_{D} - F_{c}$ for   
(upper solid curve) a driven particle with 
$q_{D} = 0.5$. The upper dashed curve is a fit 
with $\beta = 1.0$. The lower solid curve is for $q_{D} = 20$, with the
lower dashed curve a fit with $\beta = 1.47$.}
\end{figure} 

\begin{figure}
\center{
\epsfxsize=3.5in
\epsfbox{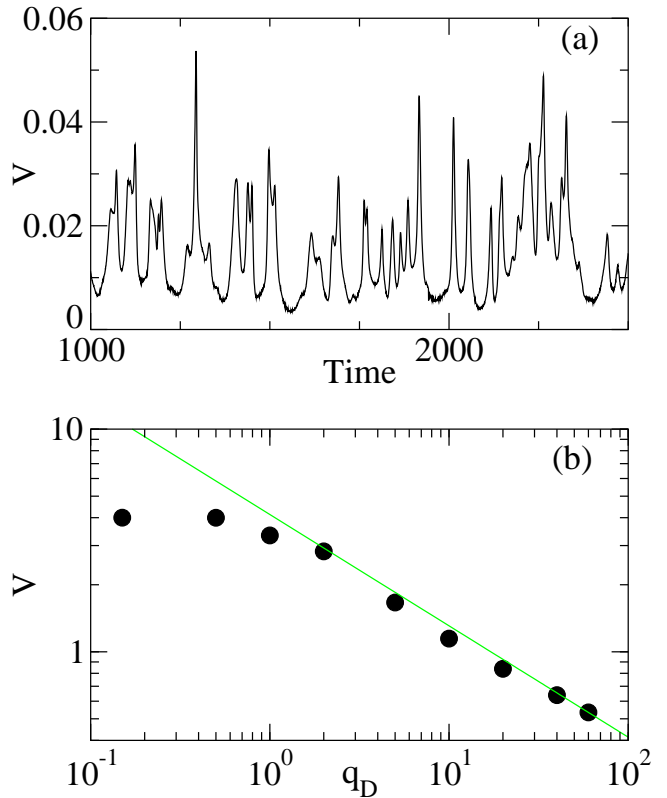}}
\caption{
(a) The velocity vs time for a fixed drive of $F_{D} = 0.75$ with
$q_{D} = 20$, showing intermittent bursts of motion. (b)
The velocity vs $q_{D}$ for a fixed drive $F_{d} = 4.0$. The
solid line is a fit of $q_{D}^{-1/2}$.}
\end{figure}  

\begin{figure}
\center{
\epsfxsize=3.5in
\epsfbox{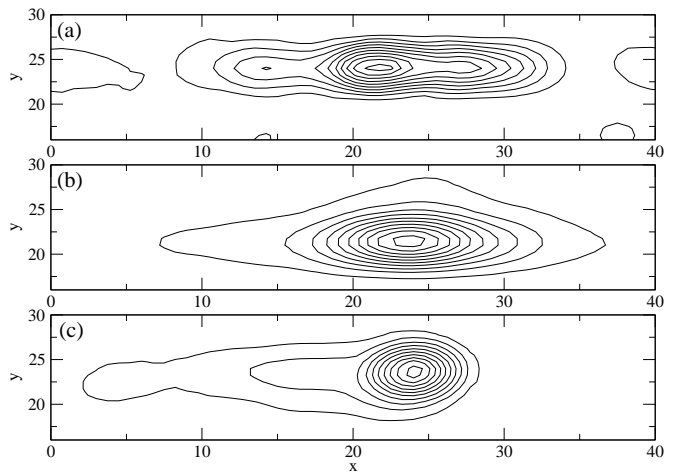}}
\caption{Contour plot of $\overline{v^2}$ 
as a function of position relative to the
driven particle.  The lowest contour is normalized to
$5\%$ of the maximum of $\overline{v^2}$,
and each contour represents a $10\%$ increase.
The system size is $48\times 48$, and the
driven particle position is fixed to $(24,24)$.}
\end{figure}

\end{document}